\documentclass[mathleft
]{an}
\usepackage{graphicx}
\usepackage{times}
\newcommand{\corot}{{\textsc{CoRoT}}}
\newcommand{\cible}{{HD~49385}}

\newcommand{\ind}[1]{_{\rm #1}}
\newcommand{\expon}[1]{^{\rm #1}}

\newcommand{\nuac}{\nu\expon{obs}\ind{ac}}
\newcommand{\vaisala}{Brunt-V\"ais\"al\"a}
\newcommand{\lsmean}{\overline{\Delta\nu}}
\def\m2s2{\,m$^{2}$\,s$^{-2}$} 
\def\cm2s1{\,cm$^{2}$\,s$^{-1}$} 

\def\aov{\alpha\ind{ov}}

\newcommand\aap{Astron. Astrophys.} 
\newcommand\solphys{Sol. Phys.} 
\newcommand\aaps{Astron. Astrophys. Suppl. Ser.} 

\begin{document}

\Pagespan{1}{}
\Yearpublication{2006}%
\Yearsubmission{2005}%
\Month{11}%
\Volume{999}%
\Issue{88}%

\title{Constraints on the core $\mu$-gradient of the solar-like star HD~49385 via low-degree mixed modes}


  
   

\author{S. Deheuvels\inst{1}\fnmsep\thanks{Corresponding author:
  \email{sebastien.deheuvels@obspm.fr}\newline}
\and  E. Michel\inst{1}
}

\titlerunning{Inference on core $\mu$-gradient via mixed modes}
\authorrunning{S. Deheuvels \& E. Michel}

\institute{
LESIA, CNRS UMR 8109,  Observatoire de Paris, Universit\'e Pierre et Marie Curie, Universit\'e Denis Diderot, 92195 Meudon cedex, France
}

\received{...}
\accepted{...}


\abstract{
The existence of an $\ell=1$ avoided crossing in the spectrum of the solar-like pulsator \corot\thanks{The CoRoT space mission, launched on 2006 December 27, was developed and
is operated by the CNES with participation of the Science Programs of ESA; ESA's RSSD, Austria, Belgium, Brazil, Germany and Spain.}-target \cible\ was established by Deheuvels \& Michel
2009. It is the first confirmed detection of such a phenomenon. The authors showed in a preliminary modeling of the star that it was in a post main
sequence status. Being a $1.3M_{\odot}$-star, \cible\ has had a convective core during its main sequence phase. 
The $\mu$-gradient left by the withdrawal of this core bears information about the processes of transport at the boundary of the core.
We here investigate the constraints that the observed avoided crossing brings on the $\mu$-gradient
in the core of the star.}

\maketitle

\section{Introduction}

Avoided crossings between non-radial oscillation modes have been the object of many theoretical studies since \cite{1974A&A....36..107S}
found that low-order p and g modes could present a mixed behavior in condensed polytropes 
In particular, \cite{1991A&A...248L..11D} stressed that the frequency at which the avoided crossing occurs has a high diagnostic potential, 
especially to constrain the processes of transport at the boundary of a convective core. \cite{2009Ap&SS.tmp..241D} have shown that low-degree avoided crossings
are associated with a distortion of the ridge of the corresponding degree. The authors explained that this feature is the effect of a coupling between $n$
modes during the phenomenon of avoided crossing and suggested that it could be used to contrain the $\mu$-gradient in the inner parts of the star.
Such a distortion was observed in the solar-like pulsator \cible, based on the analysis of 137 days of photometric data collected with the space telescope \corot\ 
(Deheuvels \& Michel 2009, \cite{2010arXiv1003.4368D}). The authors found that it could only be explained by an $\ell=1$ avoided crossing and
this enabled them to conclude that the star is in a post main sequence stage.

We here study how the observation of this avoided crossing constrains the processes of transport in the center of \cible. We first focus on the frequency at
which the avoided crossing occurs. We then show that the curvature of the $\ell=1$ ridge brings complementary information on the chemical composition at the edge of the core.

\section{Frequency at which the avoided crossing occurs \label{sect_freq_ac}}


Using the evolution code CESAM2k (\cite{1997A&AS..124..597M}) combined with the oscillation code LOSC (\cite{2008Ap&SS.316..149S}), 
we found a set of stellar models reproducing reasonably well
both the classical and the mean seismic indices of \cible.
We were then interested in selecting among these models, those which show an avoided crossing at the same frequency
as the observed avoided crossing $\nuac$. To locate the avoided crossing, we found it convenient to use the frequency of the mixed mode
which has an essentially g-mode behavior, since it varies much faster than the frequencies of the other modes. 
Based on Deheuvels et al. (2010), the fitted eigenfrequency of this mode for \cible\
is $\nuac=748.60\pm0.23\;\mu$Hz. We therefore needed to ensure that $\nuac$ corresponds to the eigenfrequency of a g mode. 

One way of doing this would be to locate the avoided crossings in each of our models. The problem is that several modes
are involved and they are not the same ones from one model to the other. It is more convenient to link the frequency of the avoided
crossing to the equilibrium quantities.
In the frame of the asymptotic theory, the eigenfrequency $\nu_{n,\ell}$ of a g mode of high order $n$ and low degree $\ell$
is approximately given by the equation
\begin{equation}
\int_{r_1}^{r_2} \left( \frac{N^2}{\nu_{n,\ell}^2}-1 \right)^{1/2} \frac{\hbox{d}r}{r} = \frac{(n-1/2)\pi}{L},
\label{eq_gmode}
\end{equation}
where 
$$L^2=\ell(\ell+1),$$
$N$ is the \vaisala\ frequency, and $r_1$, $r_2$ are the turning points of the mode in the g-mode cavity.
The integral $I(\nu)$ defined as
\begin{equation}
I(\nu) = \int_{r_1}^{r_2} \left( \frac{N^2}{\nu^2}-1 \right)^{1/2} \frac{\hbox{d}r}{r}
\end{equation}
can therefore be used to determine whether or not $\nuac$ corresponds to the eigenfrequency of a g mode.
Indeed, it is the case when there exists an order $n$ such that Eq. \ref{eq_gmode} is verified for $\nu=\nuac$. 
We will hereafter use $I(\nuac)$ as a rough indicator of how the frequency of the avoided crossing compares
in the model and in the observations. We checked a posteriori that the models which verify Eq. \ref{eq_gmode} indeed
have an avoided crossing around $\nuac$.

\subsection{Evolution of $I(\nuac)$ with the age}

\begin{figure}
\includegraphics[width=80mm]{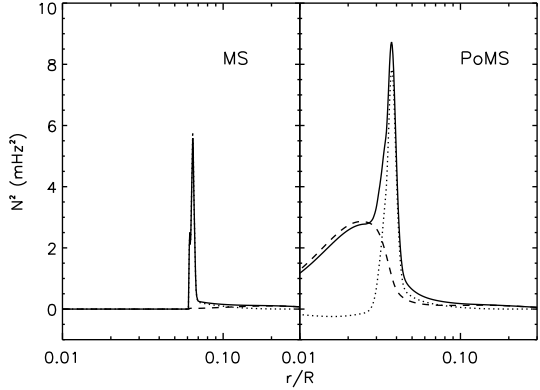}
\caption{Profile of the squared \vaisala\ frequency in the inner region of a MS and a PoMS model of \cible\ (full line). 
The dashed line shows the contribution of the departure from
adiabaticity and the dotted line, the contribution of the gradient of chemical composition.}
\label{fig_vaisala}
\end{figure}

The \vaisala\ frequency $N$ can be written as a function of the following derivatives:
\begin{equation}
\nabla\ind{ad} = \frac{\partial \ln T}{\partial \ln P} \bigg|_S \; , \; \;  \nabla = \frac{\partial \ln T}{\partial \ln P} \; , \; \;  \nabla\mu= \frac{\partial \ln \mu}{\partial \ln P},
\end{equation}
where $P$, $T$ and $\mu$ correspond to the pressure, temperature and mean molecular weight, respectively.
The subscript $S$ indicates that the definition is valid for constant entropy. It is convenient to split the expression of $N^2$ in two contributions:
\begin{equation}
N^2=\frac{g}{H_p} \left( \nabla\ind{ad} - \nabla \right) + \frac{g}{H_p} \nabla\mu ,
\label{eq_vaisala}
\end{equation}
where $H_p$ is the pressure scale height and $g$ the gravity inside the star. 
The left part of the second member in Eq. \ref{eq_vaisala} corresponds to the departure from adiabaticity. During the main sequence,
this term is null in the convective core. When the star reaches the
end of the main sequence, the nuclear reactions stop in the core, which ceases to be convective. The temperature gradient becomes
smaller than $\nabla\ind{ad}$ and the left term increases.
The right term in Eq. \ref{eq_vaisala} is caused by the gradient of chemical composition which is left by the receding convective  core.
We show in Fig. \ref{fig_vaisala} that both contributions are clearly identified both in the main sequence (MS) and in the 
post main sequence (PoMS) phase.

\begin{figure}
\includegraphics[width=80mm]{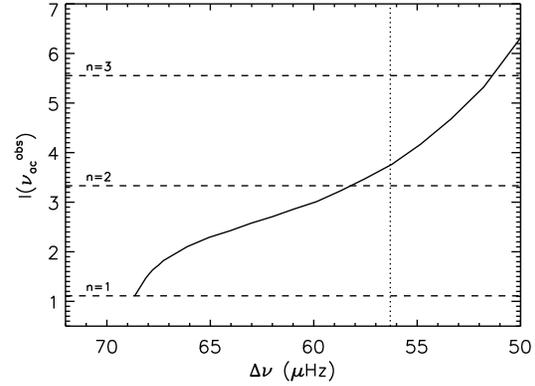}
\caption{Variations in the integral $I(\nuac)$ during the evolution of a PoMS sequence of models for \cible\ (thick solid line). The dashed lines indicate the
the solutions of Eq. \ref{eq_vaisala} for $n=1,2,3$. The dotted line represents the observed value of the mean large separation $\lsmean=56.3\,\mu$Hz.}
\label{fig_int_bv_solo}
\end{figure}

As the star evolves past the end of the main sequence, the inner regions contract to trigger hydrogen burning in shell.
Consequently the density increases in the core, which causes the factor $g/H_p$ to also increase in Eq. \ref{eq_vaisala}. 
It is therefore obvious that the integral $I(\nuac)$ increases during the evolution of the star. This can be seen for a PoMS model
of \cible\ in Fig. \ref{fig_int_bv_solo}. We note that, since the mean large separation $\lsmean$ monotonically decreases during the evolution of the star,
we used this quantity instead of the age as a parameter of evolution. This representation enables us to easily spot the models
which reproduce both the observed value of the mean large separation and the frequency of the observed avoided crossing.

\subsection{Evolution of $I(\nuac)$ with the mass}

We now consider models for which the mean large separation matches the observed one, and we study the influence of the
mass. As can be seen in Fig. \ref{fig_evol_int}, the line of the TAMS in the HR diagram becomes closer to the line of iso-$\Delta\nu$
when the mass increases. Therefore, for larger masses, the star is closer to the TAMS when it reaches the observed value of the large separation
$\lsmean\ind{obs}$ and the value of the integral $I(\nuac)$ is expected to be smaller. We indeed observe for \cible\ that the integral
decreases with increasing mass (see Fig. \ref{fig_evol_int}). With a given physics, we can find one mass for which both the value of the mean
large separation and the frequency of the avoided crossing are reproduced.

\subsection{Evolution of $I(\nuac)$ with the overshooting}

The models we here consider are in the PoMS stage. Core overshooting is not supposed to play a direct role since the convective core has vanished. However, the profile of the gradient of
chemical composition which was left behind as the convective core receded during the main sequence greatly depends on the amount of overshooting 
which existed at the edge of the core. As is usually done, we implemented overshooting in our models as an extension of the convective core over
a fraction $\aov$ of the pressure scale height $H_p$. We studied the influence of $\aov$ on the frequency of the avoided crossing.

\begin{figure}
\includegraphics[width=83mm]{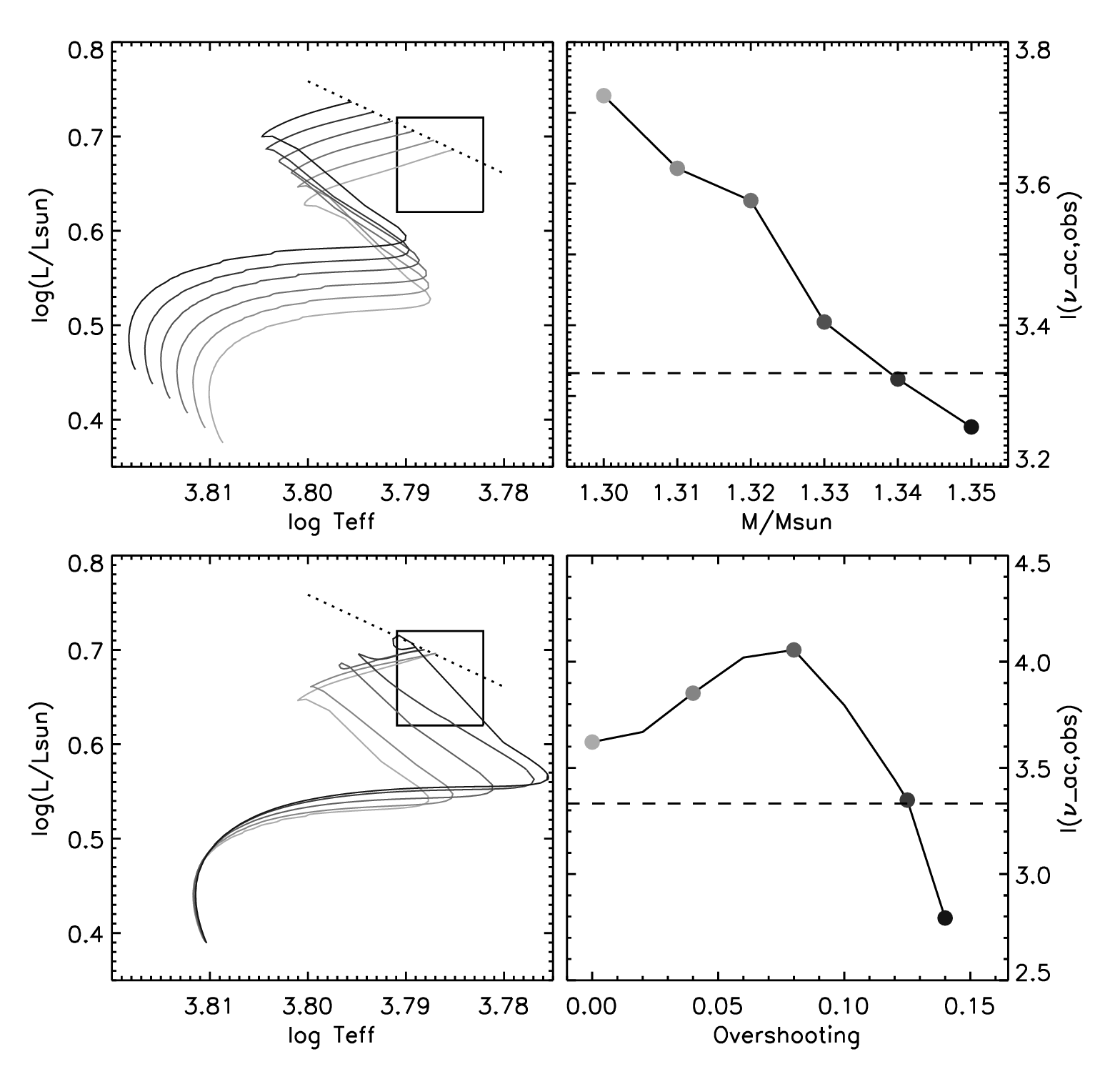}
\caption{\textbf{Top:} Evolutionary tracks in the HR diagram (left) and values of the integral $I(\nuac)$ for models of different masses fitting
$\lsmean\ind{obs}$ for \cible, with given physics ($\aov=0$). The masses range from $1.3M_{\odot}$ (light gray) to $1.36M_{\odot}$ (dark gray). In the HR diagram, the box 
represents the position of the star within 1-$\sigma$ error bars and on the dotted line, $\lsmean=\lsmean\ind{obs}$. In the right plot, the dashed line
corresponds to the solution of Eq. \ref{eq_vaisala} for $n=2$. \textbf{Bottom:} Same as the top, but for models with different $\aov$ and fixed mass ($M=1.31M_{\odot}$). The overshooting parameter ranges
from 0 (light gray) to 0.15 (dark gray).}
\label{fig_evol_int}
\end{figure}

\begin{figure}
\includegraphics[width=83mm]{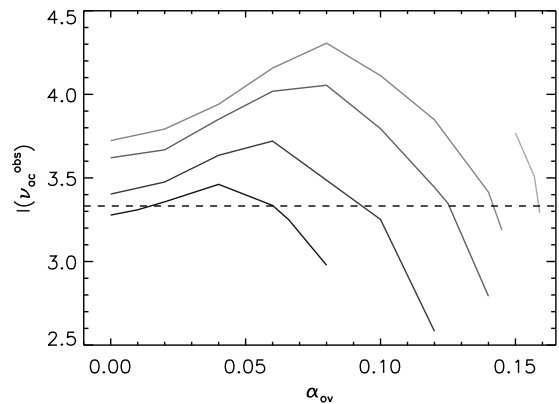}
\caption{Variations in the integral $I(\nuac)$ with overshooting for different masses. The masses range from $1.3M_{\odot}$ (light gray) to $1.36M_{\odot}$ (dark gray).
The dashed line corresponds to the solution of Eq. \ref{eq_vaisala} for $n=2$.}
\label{fig_int_os}
\end{figure}

There are two contradictory effects of core overshooting on the integral $I(\nuac)$. 
\begin{itemize}
\item First, by bringing more hydrogen to the core, the overshooting
enables the star to stay longer in the MS stage. This can be seen in Fig. \ref{fig_evol_int} in the case of \cible. Consequently, for a given mass, the bigger $\aov$,
the less the star has evolved from the TAMS when it reaches the observed value of the mean large separation. As explained above, we therefore expect
$I(\nuac)$ to decrease when $\aov$ increases. We note that above a certain value of $\aov$, the model already has a mean large separation smaller than
$\lsmean\ind{obs}$ when it reaches the TAMS, and no satisfactory model can be found. This can be used to set an upper limit on $\aov$. With the physics we use,
we found that $\aov<0.16$ for \cible. 
\item Secondly, when adding overshooting, the mixed core during the MS phase
is bigger and so the resulting He core at the TAMS is bigger. Therefore, the inner parts of the star need to be denser to start burning hydrogen in shell.
In this case, the factor $g/H_p$ in Eq. \ref{eq_vaisala} increases, and the integral $I(\nuac)$ increases. This implies that $I(\nuac)$ should increase
with $\aov$.
\end{itemize}
For a mild overshooting, the star is far from the TAMS when it reaches $\lsmean\ind{obs}$, and the first effect has little influence. We can see in Fig. \ref{fig_evol_int}
that $I(\nuac)$ increases with $\aov$ for low values of $\aov$. This tendency is reversed for large values of $\aov$.
Indeed, with a larger amount of overshooting, the star is very close to the TAMS and the first effect prevails.
Due to the competition between these two effects, certain masses lead to two different models fitting both $\lsmean\ind{obs}$ and the frequency of the
avoided crossing: one with a mild overshooting away from the TAMS, and one with a larger $\aov$, close to the TAMS. This is the case for $M=1.36M_{\odot}$
(see Fig. \ref{fig_int_os}).

In Fig. \ref{fig_int_os}, all the models which lie on the dashed line fit the position of \cible\ in the HR diagram, the observed value of the mean large separation, and 
the frequency at which the avoided crossing occurs. We found that $\aov\in[0,0.16]$ but at this stage the overshooting parameter cannot be further constrained.

\section{Curvature of the ridge}

For low-degree modes, the p-mode cavity and the g-mode cavity are strongly coupled since the evanescent zone that separates them is small.  
\cite{2009Ap&SS.tmp..241D} showed that in this case, avoided crossings involve more than two modes: for one mode trapped mainly in the
core, several consecutive-order modes trapped mainly in the envelope have a non-negligible part of their kinetic energy in the g-mode cavity.
Their frequencies are modified, resulting in a characteristic distortion of the ridge, whose shape depends on the size of the evanescent zone.
We used the curvature of the ridge as an additional constraint for our models of \cible.

\subsection{Comparison with the observations}

To efficiently compare the curvatures of the ridges, the frequency at which the avoided crossing occurs in the model must match precisely that of the observations.
The best way to ensure this is to check that the mode which is trapped mainly in the core has the same frequency in both spectra. Indeed, its frequency varies much
more rapidly than that of the other modes. Until now, we have used $I(\nuac)$ to locate the avoided crossing.
As stated above, it is only a crude indicator. Here, we made small adjustments in the age of the models selected in Sect. \ref{sect_freq_ac}
so the mode which has a dominant g-mode behavior has the frequency of the observed mode (mode $\pi_1$ in Deheuvels et al. 2010).
We then used the individual $\ell=1$ large separations to compare the curvatures of the ridges. We found that for all the satisfactory models of
Sect. \ref{sect_freq_ac}, the curvature of the $\ell=1$ ridge is less pronounced than in the observations (see Fig. \ref{fig_diag_prop}). This means that
in these models, the coupling between the g-mode cavity and the p-mode cavity is underestimated. It was pointed by \cite{1976PASJ...28..199S} that the strength of
the coupling increases when the size of the evanescent zone decreases. The size of this region needs to be decreased in the models.

The outer turning point of the g-mode cavity and the inner turning point of the p-mode cavity constitute the limits of the evanescent zone.
While the latter varies very little in the different models we considered, the former strongly depends on the $\mu$-gradient which was left
behind by the receding core during the main sequence. Indeed, we can see in Fig. \ref{fig_vaisala} that this point is located in the $\nabla\mu$
component of the \vaisala\ frequency. 

\subsection{Effect of overshooting}

\begin{figure}
\includegraphics[width=83mm]{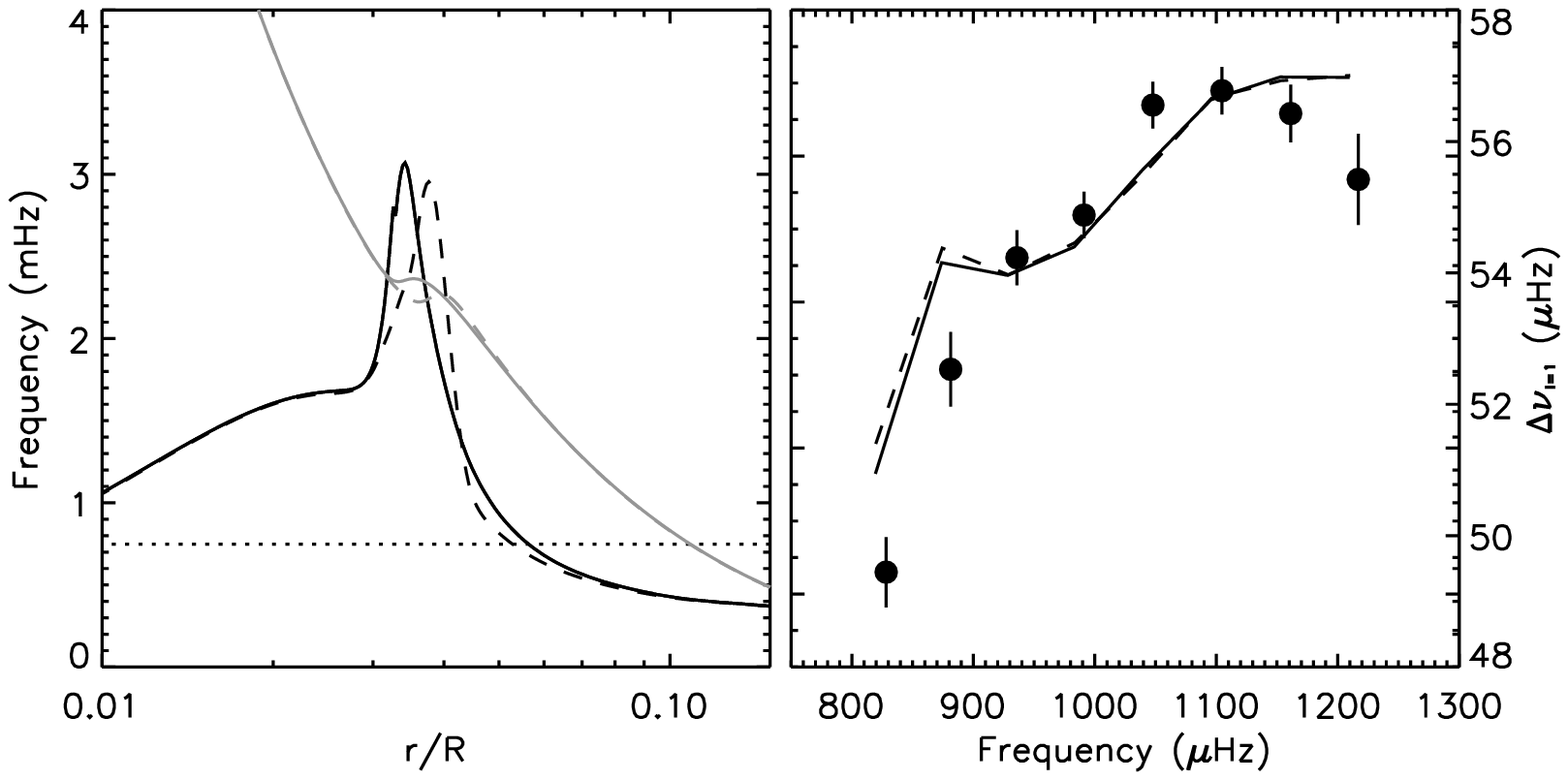}
\caption{\textbf{Left:} Propagation diagram of two models of \cible\ fitting the position
of the star in the HR diagram, the observed $\lsmean$ and the frequency of the avoided crossing: one without overshooting (full lines) 
and one with $\aov=0.7$ (dashed lines). The \vaisala\ frequency is
shown in black and the Lamb frequency for $\ell=1$ modes in gray. The dotted line indicates $\nuac$. \textbf{Right:}
Profile of the $\ell=1$ large separation for the same models. The filled circles correspond to the obervational values with
1-$\sigma$ error-bars.}
\label{fig_diag_prop}
\end{figure}

Adding core overshooting has the effect of extending the size of the mixed region
associated with the convective core and the $\mu$-gradient left by the withdrawal of the core is located less deeply in the star. 
We could therefore expect overshooting to increase the coupling, as required. However, we observe that for \cible, 
the evanescent zone is even wider with overshooting and the agreement with the observations for the curvature of the $\ell=1$ ridge
is worse (see Fig. \ref{fig_diag_prop}). 
The case of \cible\ is particular since the star is close to the TAMS. The models without overshooting
are evolved enough after the TAMS so that the nuclear reactions in shell have smoothed the $\mu$-gradient. This has the effect of reducing the size of
the evanescent zone. On the contrary, the models with overshooting reach $\lsmean\ind{obs}$
very close to the TAMS and the $\mu$-gradient is exactly the one which was left when the core receded. 
And so, eventhough the $\mu$-gradient is located less deeply, the evanescent zone is not smaller. This raises the question
of the treatment of overshooting. Here, in the overshooted region, the mixing is considered total and the gradient is supposed to be
adiabatic (known as \textit{instantaneous mixing}). Such a simple treatment of overshooting is found incompatible with the observations for \cible.

\section{Conclusion and perspectives}

Two features of the $\ell=1$ avoided crossing observed in \cible\ were reckoned to give constraints on the $\mu$-gradient in the inner parts of the
star: the frequency at which the avoided crossing occurs and the curvature of the  $\ell=1$ ridge around the avoided crossing.
By computing evolution models of \cible\ fitting the global parameters, the mean seismic indices and the frequency of the avoided
crossing, we showed that only an upper limit to the amount of overshooting that existed at the edge of the core can be set: $\alpha\ind{ov}<0.16$.
Among all these models, none is able to satisfactorily reproduce the strong curvature of the observed $\ell=1$ ridge. We conclude that
the crude description of overshooting as an instantaneous mixing which we used here (and which is most commonly used)
is incompatible with the observed spectrum of \cible. We plan to test other descriptions of the mixing which exists at the edge of the convective core.
For instance, this mixing could be treated as a diffusive process (\cite{1998A&A...334..953V}).

\acknowledgements

\newpage


\begin{thebibliography}{}
 
 \bibitem[Deheuvels \& Michel (2009)]{2009Ap&SS.tmp..241D} Deheuvels, S., \& Michel, E.\ 2009, \apss, 241 
 \bibitem[Deheuvels et al. 2010]{2010arXiv1003.4368D} Deheuvels, S., et al.\ 2010, arXiv:1003.4368 
 \bibitem[di Mauro et al. 2004]{2004SoPh..220..185D} di Mauro, M.~P., Christensen-Dalsgaard, J., Patern{\`o}, L., \& D'Antona, F.\ 2004, \solphys, 220, 185 
 \bibitem[Dziembowski \& Pamyatnykh (1991)]{1991A&A...248L..11D} Dziembowski, W.~A., \& Pamyatnykh, A.~A.\ 1991, \aap, 248, L11 
 \bibitem[Mathis et al. 2004]{2004A&A...425..243M} Mathis, S., Palacios, A., \& Zahn, J.-P.\ 2004, \aap, 425, 243 
 \bibitem[Miglio et al. 2007]{2007MNRAS.377..373M} Miglio, A., Montalb{\'a}n, J., \& Maceroni, C.\ 2007, \mnras, 377, 373 
 \bibitem[Morel 1997]{1997A&AS..124..597M} Morel, P.\ 1997, \aaps, 124, 597 
\bibitem[Osaki (1975)]{1975PASJ...27..237O} Osaki, J.\ 1975, \pasj, 27, 237 
\bibitem[Scuflaire (1974)]{1974A&A....36..107S} Scuflaire, R.\ 1974, \aap, 36, 107
 \bibitem[Scuflaire et al. 2008]{2008Ap&SS.316..149S} Scuflaire, R., Montalb{\'a}n, J., Th{\'e}ado, S., Bourge, P.-O., Miglio, A., Godart, M., Thoul, A., \& Noels, A.\ 2008, \apss, 316, 149 
 \bibitem[Shibahashi \& Osaki (1976)]{1976PASJ...28..199S} Shibahashi, H., \& Osaki, Y.\ 1976, \pasj, 28, 199 
 \bibitem[Ventura et al. 1998]{1998A&A...334..953V} Ventura, P., Zeppieri, A., Mazzitelli, I., \& D'Antona, F.\ 1998, \aap, 334, 953 

 
 \end{thebibliography}
\end{document}